\documentclass[a4paper,11pt]{article}
\pdfoutput=1

\usepackage{jheppub} 

\usepackage[T1]{fontenc} 

\RequirePackage[normalem]{ulem} 
\RequirePackage{color}\definecolor{RED}{rgb}{1,0,0}\definecolor{BLUE}{rgb}{0,0,1} 

\title{ \boldmath \quad\\[0.5cm] Search for lepton-flavor-violating tau-lepton decays to $\ell\gamma$ at Belle}

\newcounter{AffiliationCounter}
\stepcounter{AffiliationCounter}\edef\instBilbao{\protect\theAffiliationCounter}
\stepcounter{AffiliationCounter}\edef\instBonn{\protect\theAffiliationCounter}
\stepcounter{AffiliationCounter}\edef\instBNL{\protect\theAffiliationCounter}
\stepcounter{AffiliationCounter}\edef\instBINP{\protect\theAffiliationCounter}
\stepcounter{AffiliationCounter}\edef\instCharles{\protect\theAffiliationCounter}
\stepcounter{AffiliationCounter}\edef\instChonnam{\protect\theAffiliationCounter}
\stepcounter{AffiliationCounter}\edef\instCincinnati{\protect\theAffiliationCounter}
\stepcounter{AffiliationCounter}\edef\instDESY{\protect\theAffiliationCounter}
\stepcounter{AffiliationCounter}\edef\instDuke{\protect\theAffiliationCounter}
\stepcounter{AffiliationCounter}\edef\instFlorida{\protect\theAffiliationCounter}
\stepcounter{AffiliationCounter}\edef\instFuJen{\protect\theAffiliationCounter}
\stepcounter{AffiliationCounter}\edef\instFudan{\protect\theAffiliationCounter}
\stepcounter{AffiliationCounter}\edef\instGiessen{\protect\theAffiliationCounter}
\stepcounter{AffiliationCounter}\edef\instGifu{\protect\theAffiliationCounter}
\stepcounter{AffiliationCounter}\edef\instGoettingen{\protect\theAffiliationCounter}
\stepcounter{AffiliationCounter}\edef\instSokendai{\protect\theAffiliationCounter}
\stepcounter{AffiliationCounter}\edef\instGyeongsang{\protect\theAffiliationCounter}
\stepcounter{AffiliationCounter}\edef\instHanyang{\protect\theAffiliationCounter}
\stepcounter{AffiliationCounter}\edef\instHawaii{\protect\theAffiliationCounter}
\stepcounter{AffiliationCounter}\edef\instKEK{\protect\theAffiliationCounter}
\stepcounter{AffiliationCounter}\edef\instJPARC{\protect\theAffiliationCounter}
\stepcounter{AffiliationCounter}\edef\instHSE{\protect\theAffiliationCounter}
\stepcounter{AffiliationCounter}\edef\instJuelich{\protect\theAffiliationCounter}
\stepcounter{AffiliationCounter}\edef\instIKER{\protect\theAffiliationCounter}
\stepcounter{AffiliationCounter}\edef\instIISERM{\protect\theAffiliationCounter}
\stepcounter{AffiliationCounter}\edef\instIITH{\protect\theAffiliationCounter}
\stepcounter{AffiliationCounter}\edef\instIITM{\protect\theAffiliationCounter}
\stepcounter{AffiliationCounter}\edef\instIndiana{\protect\theAffiliationCounter}
\stepcounter{AffiliationCounter}\edef\instProtvino{\protect\theAffiliationCounter}
\stepcounter{AffiliationCounter}\edef\instVienna{\protect\theAffiliationCounter}
\stepcounter{AffiliationCounter}\edef\instNapoli{\protect\theAffiliationCounter}
\stepcounter{AffiliationCounter}\edef\instJAEA{\protect\theAffiliationCounter}
\stepcounter{AffiliationCounter}\edef\instJSI{\protect\theAffiliationCounter}
\stepcounter{AffiliationCounter}\edef\instKarlsruhe{\protect\theAffiliationCounter}
\stepcounter{AffiliationCounter}\edef\instKitasato{\protect\theAffiliationCounter}
\stepcounter{AffiliationCounter}\edef\instKISTI{\protect\theAffiliationCounter}
\stepcounter{AffiliationCounter}\edef\instKorea{\protect\theAffiliationCounter}
\stepcounter{AffiliationCounter}\edef\instKyotoSangyo{\protect\theAffiliationCounter}
\stepcounter{AffiliationCounter}\edef\instKyungpook{\protect\theAffiliationCounter}
\stepcounter{AffiliationCounter}\edef\instLAL{\protect\theAffiliationCounter}
\stepcounter{AffiliationCounter}\edef\instLebedev{\protect\theAffiliationCounter}
\stepcounter{AffiliationCounter}\edef\instLjubljana{\protect\theAffiliationCounter}
\stepcounter{AffiliationCounter}\edef\instLMU{\protect\theAffiliationCounter}
\stepcounter{AffiliationCounter}\edef\instMNIT{\protect\theAffiliationCounter}
\stepcounter{AffiliationCounter}\edef\instMaribor{\protect\theAffiliationCounter}
\stepcounter{AffiliationCounter}\edef\instMPI{\protect\theAffiliationCounter}
\stepcounter{AffiliationCounter}\edef\instMelbourne{\protect\theAffiliationCounter}
\stepcounter{AffiliationCounter}\edef\instMississippi{\protect\theAffiliationCounter}
\stepcounter{AffiliationCounter}\edef\instMEPhI{\protect\theAffiliationCounter}
\stepcounter{AffiliationCounter}\edef\instNagoya{\protect\theAffiliationCounter}
\stepcounter{AffiliationCounter}\edef\instNagoyaKMI{\protect\theAffiliationCounter}
\stepcounter{AffiliationCounter}\edef\instUNapoli{\protect\theAffiliationCounter}
\stepcounter{AffiliationCounter}\edef\instNara{\protect\theAffiliationCounter}
\stepcounter{AffiliationCounter}\edef\instNUU{\protect\theAffiliationCounter}
\stepcounter{AffiliationCounter}\edef\instTaiwan{\protect\theAffiliationCounter}
\stepcounter{AffiliationCounter}\edef\instKrakow{\protect\theAffiliationCounter}
\stepcounter{AffiliationCounter}\edef\instNiigata{\protect\theAffiliationCounter}
\stepcounter{AffiliationCounter}\edef\instNovosibirsk{\protect\theAffiliationCounter}
\stepcounter{AffiliationCounter}\edef\instOsakaCity{\protect\theAffiliationCounter}
\stepcounter{AffiliationCounter}\edef\instPNNL{\protect\theAffiliationCounter}
\stepcounter{AffiliationCounter}\edef\instPanjab{\protect\theAffiliationCounter}
\stepcounter{AffiliationCounter}\edef\instPeking{\protect\theAffiliationCounter}
\stepcounter{AffiliationCounter}\edef\instPittsburgh{\protect\theAffiliationCounter}
\stepcounter{AffiliationCounter}\edef\instPunjab{\protect\theAffiliationCounter}
\stepcounter{AffiliationCounter}\edef\instNPC{\protect\theAffiliationCounter}
\stepcounter{AffiliationCounter}\edef\instRIKENMSL{\protect\theAffiliationCounter}
\stepcounter{AffiliationCounter}\edef\instUSTC{\protect\theAffiliationCounter}
\stepcounter{AffiliationCounter}\edef\instSeoul{\protect\theAffiliationCounter}
\stepcounter{AffiliationCounter}\edef\instShoyaku{\protect\theAffiliationCounter}
\stepcounter{AffiliationCounter}\edef\instSoongsil{\protect\theAffiliationCounter}
\stepcounter{AffiliationCounter}\edef\instSungkyunkwan{\protect\theAffiliationCounter}
\stepcounter{AffiliationCounter}\edef\instSydney{\protect\theAffiliationCounter}
\stepcounter{AffiliationCounter}\edef\instTabuk{\protect\theAffiliationCounter}
\stepcounter{AffiliationCounter}\edef\instTata{\protect\theAffiliationCounter}
\stepcounter{AffiliationCounter}\edef\instTUM{\protect\theAffiliationCounter}
\stepcounter{AffiliationCounter}\edef\instToho{\protect\theAffiliationCounter}
\stepcounter{AffiliationCounter}\edef\instTohoku{\protect\theAffiliationCounter}
\stepcounter{AffiliationCounter}\edef\instERI{\protect\theAffiliationCounter}
\stepcounter{AffiliationCounter}\edef\instTokyo{\protect\theAffiliationCounter}
\stepcounter{AffiliationCounter}\edef\instTIT{\protect\theAffiliationCounter}
\stepcounter{AffiliationCounter}\edef\instTMU{\protect\theAffiliationCounter}
\stepcounter{AffiliationCounter}\edef\instVPI{\protect\theAffiliationCounter}
\stepcounter{AffiliationCounter}\edef\instWayneState{\protect\theAffiliationCounter}
\stepcounter{AffiliationCounter}\edef\instYamagata{\protect\theAffiliationCounter}
\stepcounter{AffiliationCounter}\edef\instYonsei{\protect\theAffiliationCounter}

\collaborationImg{\includegraphics[width=2cm]{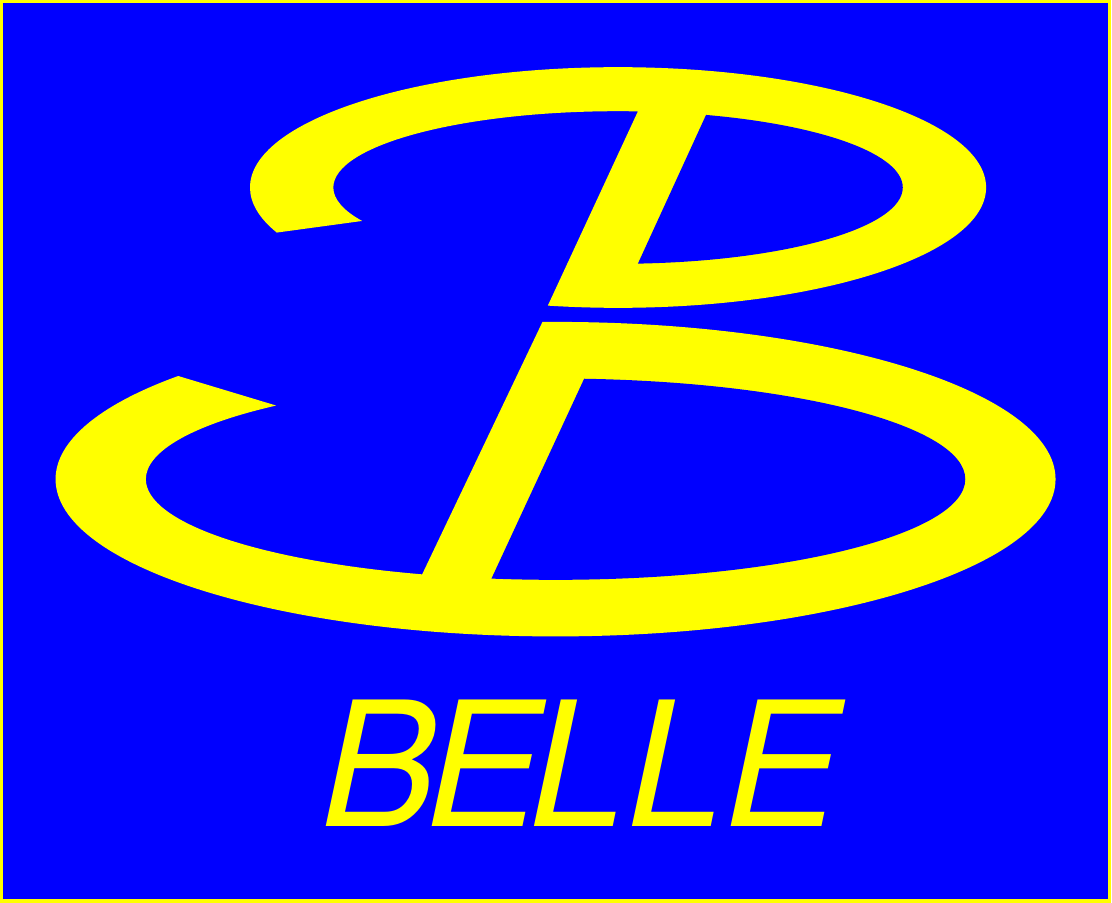}}
\collaboration{The Belle Collaboration}
  \author[\instNiigata]{K.~Uno\note[$\dagger$]{Corresponding author},} 
  \author[\instNiigata]{K.~Hayasaka,} 
  \author[\instNagoya]{K.~Inami,} 
  \author[\instKEK,\instSokendai]{I.~Adachi,} 
  \author[\instTokyo]{H.~Aihara,} 
  \author[\instBNL]{D.~M.~Asner,} 
  \author[\instCincinnati]{H.~Atmacan,} 
  \author[\instHSE]{T.~Aushev,} 
  \author[\instTabuk]{R.~Ayad,} 
  \author[\instDESY]{V.~Babu,} 
% ...

\emailAdd{uno@hep.sc.niigata-u.ac.jp}

\abstract{
Charged lepton flavor violation is forbidden in the Standard Model but possible in several new physics scenarios.
  In many of these models, the radiative decays $\tau^{\pm}\rightarrow\ell^{\pm}\gamma$~($\ell=e,\mu$) are predicted to have a sizeable probability, making them particularly interesting channels to search at various experiments.
  An updated search via $\tau^{\pm}\rightarrow\ell^{\pm}\gamma$ using full data of the Belle experiment, corresponding to an integrated luminosity of 988 fb$^{-1}$, is reported for charged lepton flavor violation.
  No significant excess over background predictions from the Standard Model is observed, and the upper limits on the branching fractions, $\mathcal{B}(\tau^{\pm}\rightarrow \mu^{\pm}\gamma)$ $\leq$ $4.2\times10^{-8}$ and $\mathcal{B}(\tau^{\pm}\rightarrow e^{\pm}\gamma)$ $\leq$ $5.6\times10^{-8}$, are set at 90\% confidence level.
}

\keywords{Taus, Lepton number, Charged lepton flavor violation}

\preprint{\vbox{ \hbox{   }
\hbox{Belle Preprint 2021-09}
\hbox{KEK Preprint 2021-5}
}}

\begin{document} 
\maketitle
\flushbottom
\section{Introduction}
Charged lepton flavor violation~(CLFV) is forbidden in the Standard Model but occurs with a yet unobservably small probability, $\mathcal{O}$(10$^{-40}$), via neutrino oscillations~\cite{SMnu}.
However, it is enhanced in theories beyond the Standard Model~(BSM) such as Minimal Supersymmetric Standard Model, grand unified theories and seesaw mechanisms~\cite{MSSM,GUT,SeeSaw}.
Several BSM models predict CLFV processes occurring at an observable level in experiments.
An observation of CLFV would be a clear signature of BSM, making the search for this phenomenon one of the high-priority physics tasks.

In several models~\cite{MSSM,GUT,SeeSaw}, the radiative decays $\tau^{\pm}\rightarrow\ell^{\pm}\gamma$~($\ell=e,\mu$) have a sizeable probability, making them highly motivated channels.
In the past, searches for $\tau^{\pm}\rightarrow\ell^{\pm}\gamma$ were performed by the Belle and BaBar experiments~\cite{HAYASAKA200816,BhaBhaResult}.
Belle used 535~fb$^{-1}$ data corresponding to $477\times10^{6}$ tau pairs~($N_{\tau\tau}$) delivered by the KEKB asymmetric-energy $e^{+}e^{-}$ collider~\cite{KEKB} and set upper limits on the branching fractions at the 90\% confidence level:
$\mathcal{B}(\tau^{\pm}\rightarrow\mu^{\pm}\gamma)<4.5\times10^{-8} $ and $\mathcal{B}(\tau^{\pm}\rightarrow e^{\pm}\gamma) < 1.2\times10^{-7}$~\cite{HAYASAKA200816}.
Similarly, BaBar set upper limits by using 516~fb$^{-1}$ data equivalent to $N_{\tau\tau}=480\times10^{6}$ delivered by the PEP-II asymmetric-energy $e^{+}e^{-}$ collider~\cite{PEP2}: $\mathcal{B}(\tau^{\pm}\rightarrow\mu^{\pm}\gamma)<4.4\times10^{-8} $ and $\mathcal{B}(\tau^{\pm}\rightarrow e^{\pm}\gamma) < 3.3\times10^{-8}$~\cite{BhaBhaResult}.

In this paper, an update search for $\tau^{\pm}\rightarrow\ell^{\pm}\gamma$ decays at the Belle experiment is reported.
Since the tau pairs are produced via the $e^{+}e^{-}\rightarrow\tau^{+}\tau^{-}$ process, we use all $\Upsilon(nS)$ resonance data corresponding to a luminosity of 5.7~fb$^{-1}$ at $\Upsilon(1S)$, 24.9~fb$^{-1}$ at $\Upsilon(2S)$, 2.9~fb$^{-1}$ at $\Upsilon(3S)$, 711~fb$^{-1}$ at $\Upsilon(4S)$, and 121.4~fb$^{-1}$ at the $\Upsilon(5S)$ resonance~\cite{Luminosity}.
In addition, a data sample recorded 60~MeV below the $\Upsilon(4S)$ resonance is used~\cite{Luminosity}.
The total integrated luminosity is 988 fb$^{-1}$, which corresponds to $N_{\tau\tau}=912\times10^{6}$~\cite{Luminosity}. This sample represents the largest number of tau-pair events recorded by a single $e^{+}e^{-}$ experiment.

The Belle detector was a large-solid-angle magnetic spectrometer consisting of a silicon vertex detector~(SVD), a 50-layer central drift chamber~(CDC), an array of aerogel threshold Cherenkov counters~(ACC), a barrel-like arrangement of time-of-flight scintillation counters~(TOF), and an electromagnetic calorimeter~(ECL) comprising CsI(Tl) crystals. All these components are located inside a superconducting solenoid coil that provides a 1.5~T magnetic field.  An iron flux-return located outside of the coil is instrumented with resistive plate chambers to detect $K_L^0$ mesons and muons~(KLM).
The detector is described in detail elsewhere~\cite{Belle}.

This analysis uses Monte Carlo~(MC) simulated samples to optimize event selection as well as to estimate signal and background contributions.
Signal MC samples and generic $\tau^{+}\tau^{-}$ processes are generated by KKMC and TAUOLA~\cite{KKMC}.
Other background processes, namely, $e^{+}e^{-}\gamma$~($e^{+}e^{-}\rightarrow e^{+}e^{-}\gamma$), $\mu^{+}\mu^{-}\gamma$~($e^{+}e^{-}\rightarrow\mu^{+}\mu^{-}\gamma$), two-photon~($e^{+}e^{-}\rightarrow e^{+}e^{-}\ell^{+}\ell^{-}$), and $q\bar{q}~(e^{+}e^{-}\rightarrow q\bar{q},~q=u,d,s,c,b)$ events are generated by BHLUMI~\cite{BHLUMI}, KKMC~\cite{KKMC}, AAFH~\cite{AAFHB}, and EvtGen~\cite{EvtGen}, respectively.
Signal MC samples are $\tau^{+}\tau^{-}$ pair events with one of the taus decaying to the $\ell^{\pm}\gamma$ final state and the other generically.
The detector simulation is done using GEANT3-based program~\cite{GEANT3}.

\section{Event selection}
\begin{figure}[htbp]
  \begin{center}
    \includegraphics[width=70mm]{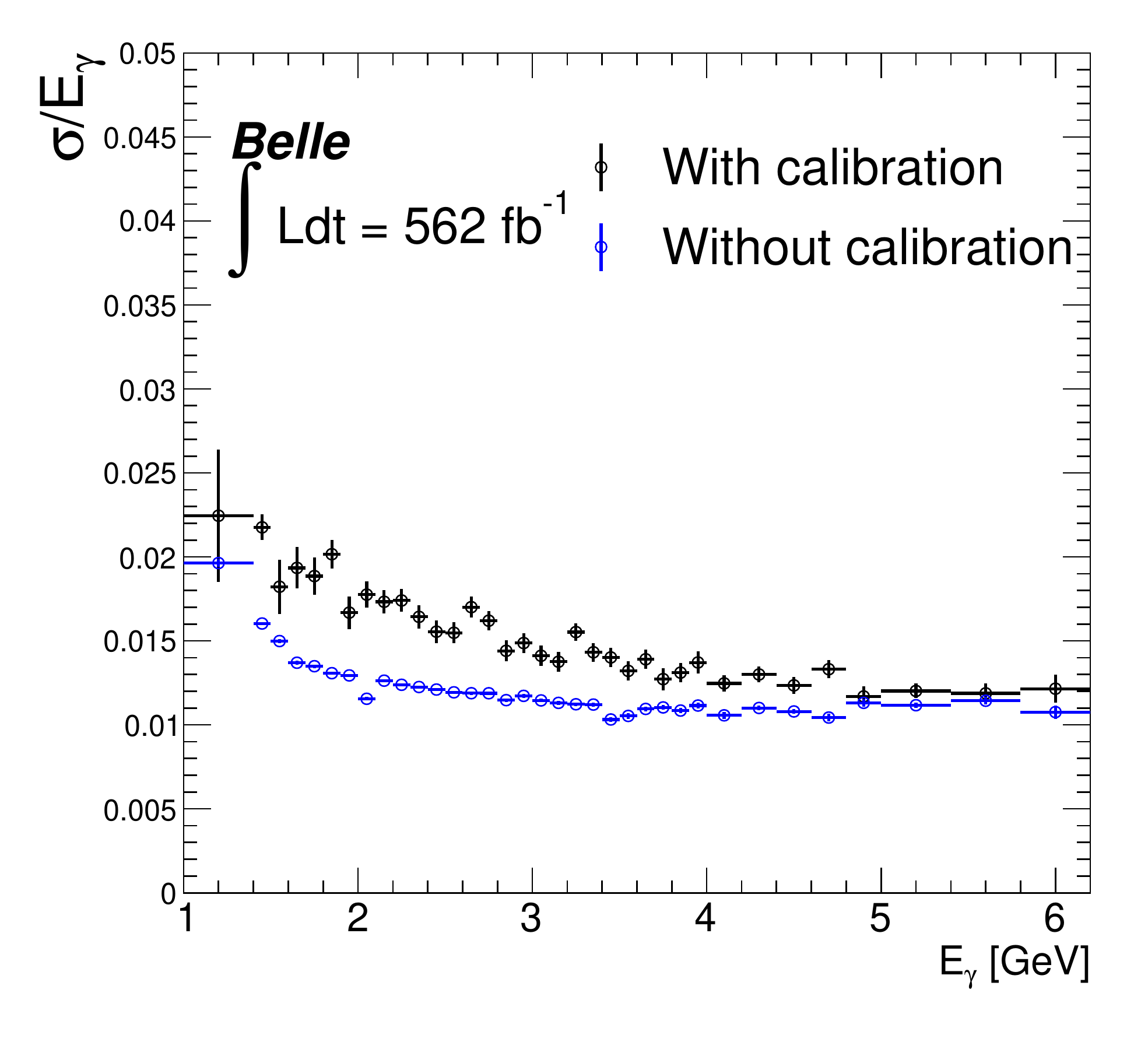}
  \end{center}
\caption{ \label{fig:photon_reso} 
  Energy resolution as a function of the reconstructed photon energy in the $e^{+}e^{-}\rightarrow\mu^{+}\mu^{-}\gamma$ events.
  Black~(Blue) points are the photon energy resolution with~(without) the calibration applied.
  Error bars are the statistical uncertainties.
}
\end{figure}

Photon candidates are selected from ECL clusters that are consistent with an electromagnetic shower but not associated with any charged tracks.
This analysis uses a photon with energy from 100 MeV to 6 GeV, and is thus sensitive to the photon energy resolution over a broad energy range.
We have revised the photon-energy calibration method using the $e^{+}e^{-}\rightarrow\mu^{+}\mu^{-}\gamma$ events for the first time at Belle.
The photon energy resolution is evaluated by subtracting the recoil energy of the $\mu^{+}\mu^{-}$ system from the photon energy measured in the ECL for data and MC simulation.
Figure~\ref{fig:photon_reso} shows the energy resolution obtained as a function of the reconstructed photon energy in the $e^{+}e^{-}\rightarrow\mu^{+}\mu^{-}\gamma$ events.
The calibrated resolution in simulation agrees with that in data as well as is compatible with the test-beam result~\cite{TestBeam}.
This is a major improvement with respect to the previous analysis~\cite{HAYASAKA200816}.

Muon candidates are identified using a likelihood ratio, $\mathcal{L_{\mu}}$, which is based on the difference between the range of the track calculated from the particle momentum and that measured in the KLM.
This ratio includes the value of $\chi^{2}$ formed from the KLM hit locations with respect to the extrapolated track.
The muon identification efficiency for the selection applied $\mathcal{L_{\mu}}>0.95$ is 90\%, with a pion misidentification probability of 0.8\%~\cite{Muon}.
Identification of electrons uses an analogous likelihood ratio, $\mathcal{L}_{e}$, based on specific ionization from the CDC, the ratio of the energy deposited in the ECL to the momentum measured by the CDC and SVD combined,
the shower shape in the ECL, hit information from the ACC, and matching between the position of the charged track and the ECL cluster.
The electron identification efficiency for the selection applied $\mathcal{L}_{e}>0.9$ is 95\%, with a pion misidentification probability of 0.07\%~\cite{Electron}.

We follow a blind analysis approach in this search, where the data in the interesting kinematic region remain hidden until the selection criteria and background estimation strategy are finalized.
All selection criteria are optimized in order to maximize the search sensitivity, $\epsilon/\sqrt{N_{\mathrm{bkg}}}$, where $\epsilon$ is the overall signal efficiency and $N_{\mathrm{bkg}}$ is the number of background events.
Since we use all $\Upsilon(nS)$ resonance data with different center-of-mass energy $\sqrt{s}$, some of the selection variables are scaled by $\sqrt{s}$.

The following preselection criteria are applied in this search.
Exactly two oppositely charged track are required to make the event's net charge zero to suppress $q\bar{q}$ events.
Candidate events are retained if both tracks have $p^{\mathrm{CM}}\leq0.43\sqrt{s}$ GeV/$c$, $p_{\mathrm{T}}\geq0.1$ GeV/$c$ and $-0.866<\cos\theta_{\mathrm{track}}<0.956$ in order to reduce $e^{+}e^{-}\gamma$, $\mu^{+}\mu^{-}\gamma$, and two-photon events
Here, $\theta_{\mathrm{track}}$ is the polar angle of the track in the laboratory frame.
For the search of $\tau^{\pm}\rightarrow e^{\pm}\gamma$ decays, the tracks that go through gaps between ECL crystals must be rejected to avoid misidentification of electrons.
Thus, the tracks are required to lie within the ECL acceptance, $\cos\theta_{\mathrm{track}}\in[-0.907,-0.652]\cup[-0.602,0.829]\cup[0.854,0.956]$.
Photons are required to have an energy $E_{\gamma}>0.1$~GeV within the region, $-0.625<\cos\theta_{\mathrm{\gamma}}<0.846$, where $\theta_{\gamma}$ is the polar angle of the photon in the laboratory frame.

A $\tau^{+}\tau^{-}$ pair event is divided into two hemispheres in the CM frame using a thrust vector~\cite{thrust}: signal- and tag-side tau.
The signal-side tau decays to a muon~(electron) and a photon for the $\tau^{\pm}\rightarrow\mu^{\pm}\gamma$~($\tau^{\pm}\rightarrow e^{\pm}\gamma$) search.
The number of photons in the signal side should be exactly one, which must have $E_{\gamma}>0.5$ GeV and $-0.602<\cos\theta_{\gamma}<0.829$ to suppress misreconstructed photons.

The tag-side tau is assumed to undergo one-prong decays such as $\tau\rightarrow e\nu\bar{\nu}$, $\mu\nu\bar{\nu}$, $\pi\nu$, and $\rho\nu$.
If the track in the tag side is identified as an electron or a muon, the event is classified as a leptonic channel.
Otherwise, the event is classified as a $\pi$ or $\rho$ channel.
If there are no photons in the tag side, the event is classified as a $\pi$ channel. Otherwise, it is a $\rho$ channel.
In order to reduce the $\mu^{+}\mu^{-}\gamma$~($e^{+}e^{-}\gamma$) contamination, an extra muon~(electron) is vetoed using the criterion, $\mathcal{L_{\mu}}<0.1$~($\mathcal{L}_{e}<0.1$) for $\tau^{\pm}\rightarrow\mu^{\pm}\gamma$~($\tau^{\pm}\rightarrow e^{\pm}\gamma$) search.

After preselecting events, the following selection criteria are applied to further suppress background events.
The total visible energy in the CM frame, $E_{\mathrm{total}}^{\mathrm{CM}}/\sqrt{s}$, is required to be smaller than 0.93 for the leptonic channel, 0.86 for the $\pi$ channel, and 0.94 for the $\rho$ channel.
Since the energy of neutrinos is different for these channels, the quantitative criteria are accordingly changed for them.
For the $\rho$ channel, an energy sum of the two charged tracks and the photon in the signal side, $E_{\mathrm{sum}}^{\mathrm{CM}}/\sqrt{s}$, is also required to be smaller than 0.86 due to extra $\pi^{0}$ in the tag side, while no such requirement is applied for other channels.
These requirements further suppress the $e^{+}e^{-}\gamma$ and $\mu^{+}\mu^{-}\gamma$ events.
The cosine of the angle between the two tracks, $\cos\theta_{\mathrm{track(sig, tag)}}$, and that between the track and the photon in the signal side, $\cos\theta_{\ell\gamma}$, are required to be $\cos\theta_{\mathrm{track(sig, tag)}}<0.0$, and $0.4 <\cos\theta_{\ell\gamma}<0.8$, respectively, to reject $\tau^{+}\tau^{-}$ background events that contain $\pi^{0}$'s from tau decays.

\begin{figure}[htbp]
 \begin{minipage}{0.45\hsize}
  \begin{center}
   \includegraphics[width=70mm]{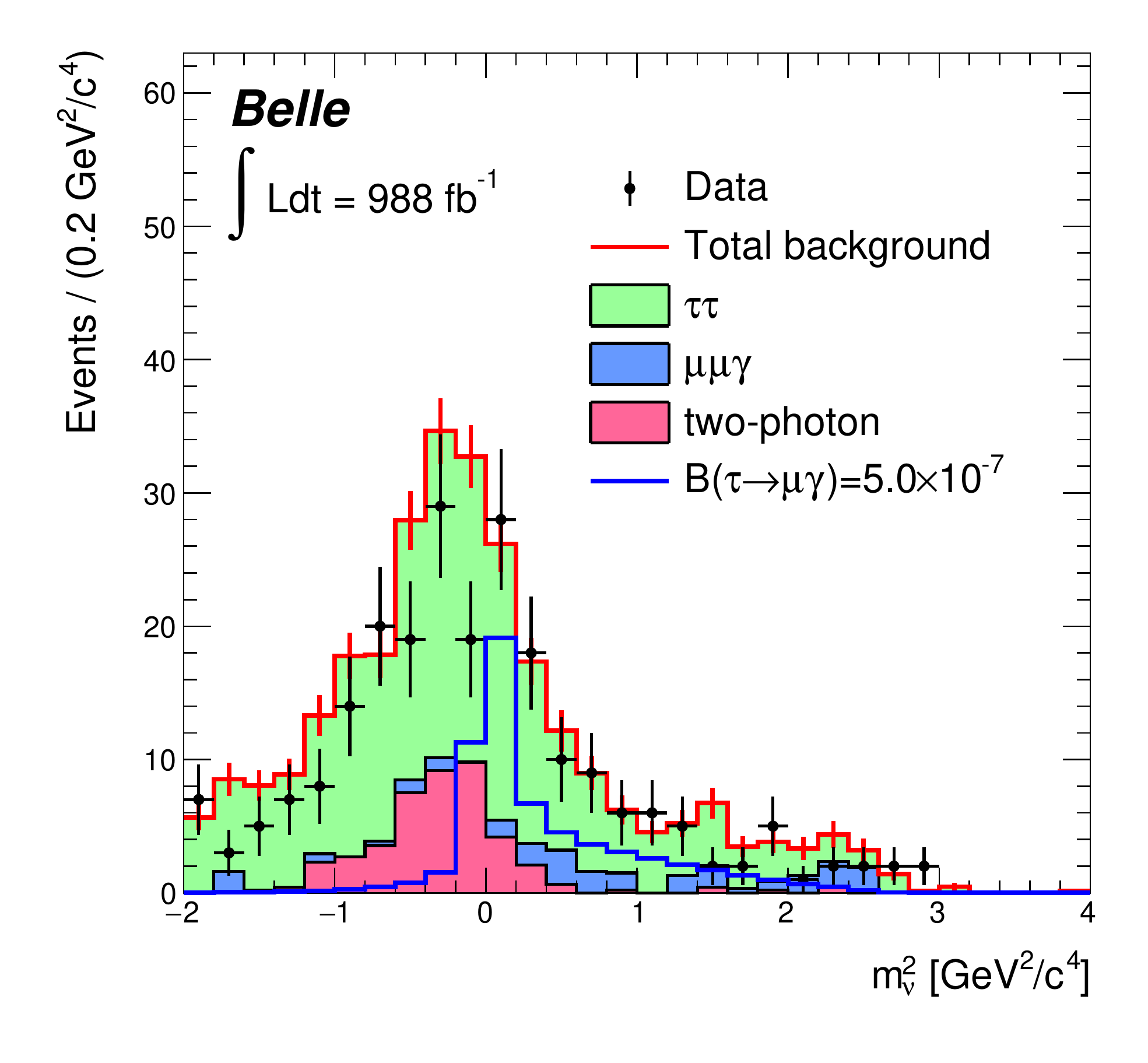}
   \hspace{1.5 cm} {(a) $\tau^{\pm}\rightarrow\mu^{\pm}\gamma$}
  \end{center}
 \end{minipage}
 \begin{minipage}{0.45\hsize}
  \begin{center}
   \includegraphics[width=70mm]{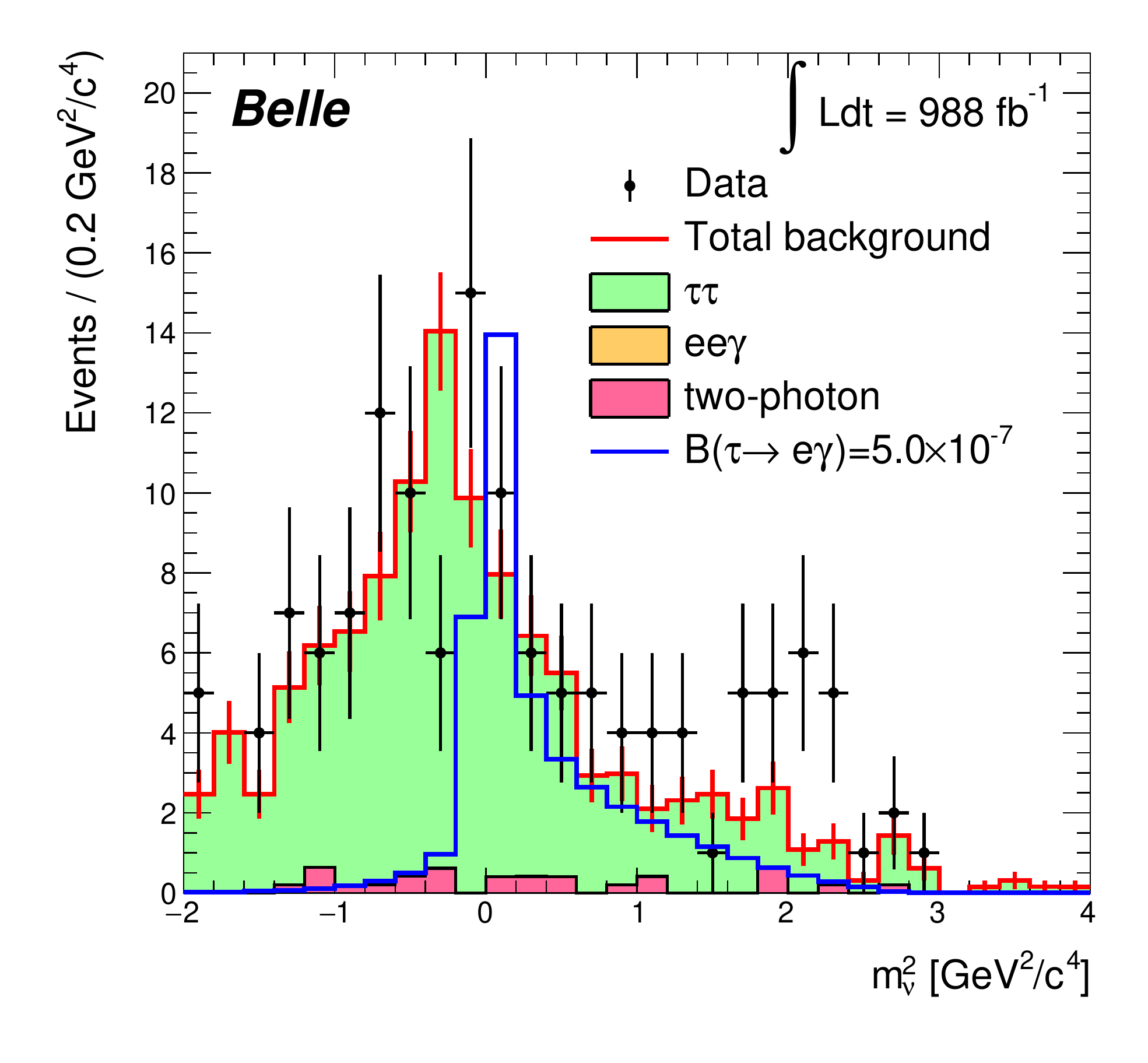}
   \hspace{1.5 cm} {(b) $\tau^{\pm}\rightarrow e^{\pm}\gamma$}
  \end{center}
 \end{minipage}
\caption{ \label{fig:sr_mnu2} 
  Distribution of missing-mass-squared on the tag side~($m_{\nu}^{2}$) for (a)~$\tau^{\pm}\rightarrow\mu^{\pm}\gamma$ and (b)~$\tau^{\pm}\rightarrow e^{\pm}\gamma$ channels.
  Events satisfying all selection criteria except for the $m_{\nu}^{2}$ requirement and $M_{\mathrm{bc}}\in[1.73,~1.85]$ GeV/$c^{2}$ are plotted.
  The background MC samples are normalized to the cross section times integrated luminosity of 988 fb$^{-1}$.
  The blue histograms show the signal MC samples with an assumed branching fraction $\mathcal{B}(\tau^{\pm}\rightarrow\ell^{\pm}\gamma)=5.0\times10^{-7}$.
}
\end{figure}

The missing momentum is calculated by subtracting the sum of the three-momenta of all charged tracks and photons from the sum of the beam momenta in laboratory frame.
Its magnitude $|\vec{p}_{\mathrm{miss}}|$ is required to be greater than $0.4$ GeV/$c$.
The cosine of the polar angle of $\vec{p}_{\mathrm{miss}}$ is required to be $-0.866 < \cos\theta_{\mathrm{miss}}<0.956$.
A criterion on the cosine of the angle between $\vec{p}_{\mathrm{miss}}$ and the tag-side track, $0.4 < \cos\theta_{\mathrm{miss,track(tag)}}< 0.98$~($0.4 < \cos\theta_{\mathrm{miss,track(tag)}}< 0.99$) for $\tau^{\pm}\rightarrow\mu^{\pm}\gamma$~($\tau^{\pm}\rightarrow e^{\pm}\gamma$) search is also required.
These requirements can suppress $e^{+}e^{-}\gamma$ and $\mu^{+}\mu^{-}\gamma$ events.
We define the missing-mass-squared on the tag side as $m_{\nu}^{2} = (E_{\ell\gamma}^{\mathrm{CM}} - E_{\mathrm{tag}}^{\mathrm{CM}})^{2} - |\vec{p}_{\mathrm{miss}}^{\mathrm{~CM}}|^{2}$, where $E_{\ell\gamma}^{\mathrm{CM}}$~($E_{\mathrm{tag}}^{\mathrm{CM}}$) is the sum of the energy of the signal~(tag) side in the CM frame, to reduce background events. Here, the natural unit $c=1$ is used in the formula throughout the paper.
Since $\tau^{+}\tau^{-}$ events are produced back-to-back in the CM frame and there are no neutrinos in the signal side for $\tau^{\pm}\rightarrow\ell^{\pm}\gamma$ events, the energy of tag-side tau in the CM frame is taken as that of the signal-side tau, $E_{\ell\gamma}^{\mathrm{CM}}$ and the missing momentum of tag-side tau is taken as that of the whole event.
Figure~\ref{fig:sr_mnu2} shows the distribution of $m_{\nu}^{2}$. The signal distribution is distinct from background due to the kinematic difference.
Since the distribution depends on the number of neutrinos, a quantitative criterion is accordingly adjusted for each channel; the specific requirements are $0.0~\mathrm{GeV}^{2}$/$c^{4}<m_{\nu}^{2}<2.8~\mathrm{GeV}^{2}$/$c^{4}$ for the leptonic channel, $-0.1$~GeV$^{2}$/$c^{4}$ $<m_{\nu}^{2}<1.2$~GeV$^{2}$/$c^{4}$ for the $\pi$ channel, and $-0.3$~GeV$^{2}$/$c^{4}$ $<m_{\nu}^{2}<1.5$~GeV$^{2}$/$c^{4}$ for the $\rho$ channel in order to reduce $\tau^{+}\tau^{-}$ background events.

\begin{figure}[htbp]
 \begin{minipage}{0.45\hsize}
  \begin{center}
   \includegraphics[width=70mm]{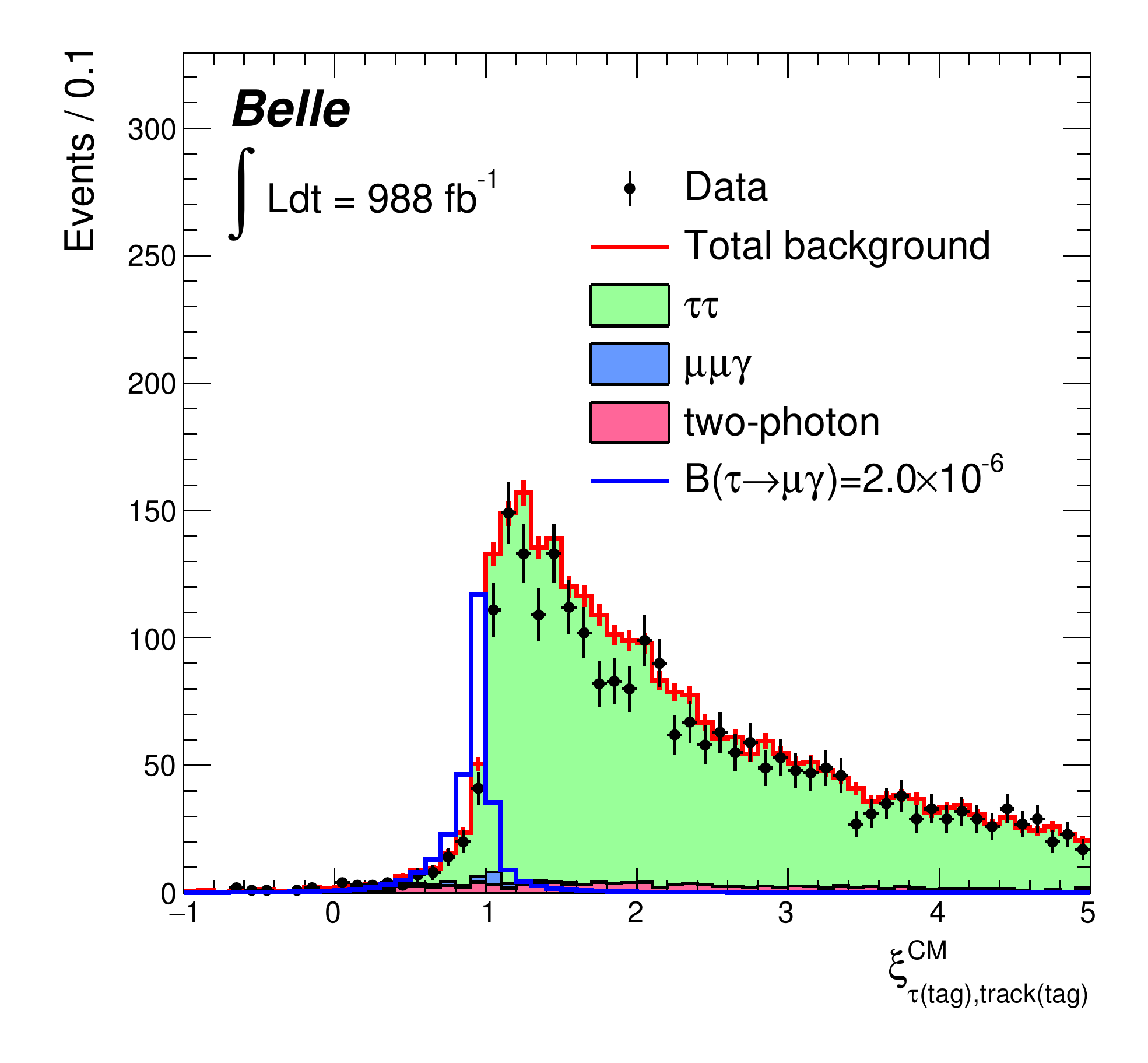}
   \hspace{1.5 cm} {(a) $\tau^{\pm}\rightarrow\mu^{\pm}\gamma$}
  \end{center}
 \end{minipage}
 \begin{minipage}{0.45\hsize}
  \begin{center}
   \includegraphics[width=70mm]{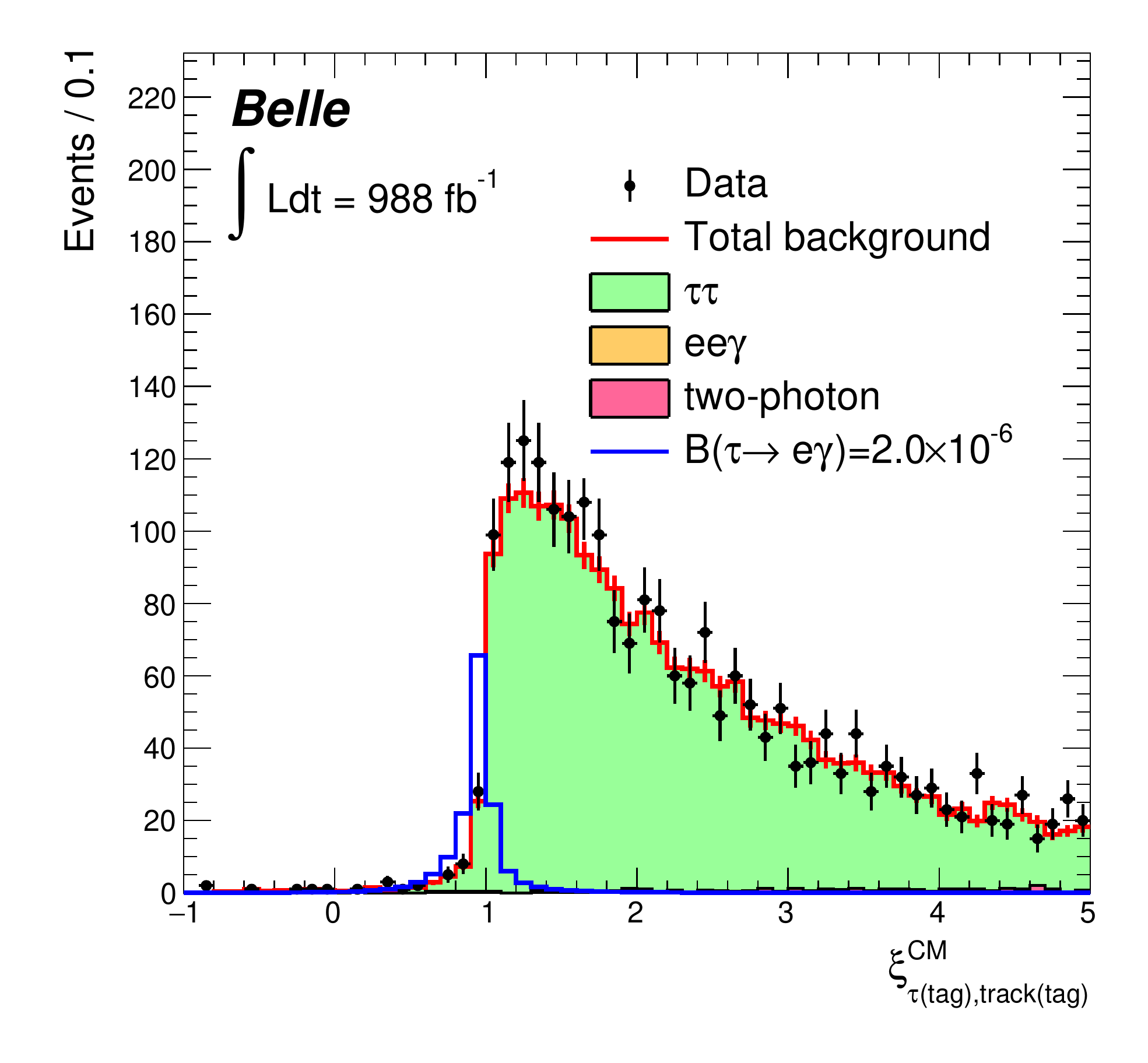}
   \hspace{1.5 cm} {(b) $\tau^{\pm}\rightarrow e^{\pm}\gamma$}
  \end{center}
 \end{minipage}
\caption{ \label{fig:sr_cos_exp} 
  Distribution of $\xi_{\tau(\mathrm{tag}),\mathrm{track(tag)}}^{\mathrm{~CM}}$ for (a)~$\tau^{\pm}\rightarrow\mu^{\pm}\gamma$ and (b)~$\tau^{\pm}\rightarrow e^{\pm}\gamma$ channels.
  Events satisfying all selection criteria except for the $\xi_{\tau(\mathrm{tag}),\mathrm{track(tag)}}^{\mathrm{~CM}}$ requirement and $M_{\mathrm{bc}}\in[1.73,~1.85]$ GeV/$c^{2}$ are plotted. 
 The background MC samples are normalized to the cross section times integrated luminosity of 988 fb$^{-1}$.
 The blue histograms show the signal MC samples with an assumed branching fractions $\mathcal{B}(\tau^{\pm}\rightarrow\ell^{\pm}\gamma)=2.0\times10^{-6}$.
}
\end{figure}

In order to improve search sensitivity, two more variables are introduced.
The first one is an energy asymmetry between the lepton and the photon in the signal side, $|E_{\ell}^{\mathrm{CM}}-E_{\gamma}^{\mathrm{CM}}|/(E_{\ell}^{\mathrm{CM}}+E_{\gamma}^{\mathrm{CM}})$.
The signal events are two-body decays, while the main background arises from three-body decays, $\tau^{\pm}\rightarrow\ell^{\pm}\nu_{\ell}\nu_{\tau}$. Thus, the energy asymmetry should be larger in background events.
We apply a requirement of $|E_{\ell}^{\mathrm{CM}}-E_{\gamma}^{\mathrm{CM}}|/(E_{\ell}^{\mathrm{CM}}+E_{\gamma}^{\mathrm{CM}})<0.65$.
The second variable, $\xi_{\tau(\mathrm{tag}),\mathrm{track(tag)}}^{\mathrm{~CM}}$ is defined as follows.
The missing mass squared against a charged track in the tag-side tau is written as
\begin{eqnarray}
  m_{\mathrm{miss.track(tag)}}^{2} &=& [p_{\tau(\mathrm{tag})}^{\mathrm{CM}}-p_{\mathrm{track(tag)}}^{\mathrm{CM}}]^{2} \\ \nonumber
  &=& m_{\tau(\mathrm{tag})}^{2} + m_{\mathrm{track(tag)}}^{2} -2[ E_{\tau(\mathrm{tag})}^{\mathrm{CM}} E_{\mathrm{track(tag)}}^{\mathrm{CM}} - \vec{p}^{~\mathrm{CM}}_{\tau(\mathrm{tag})}\cdot\vec{p}_{\mathrm{track(tag)}}^{~\mathrm{CM}}],  \nonumber
\end{eqnarray}
where $p_{\tau(\mathrm{tag})}^{\mathrm{CM}}=[E_{\tau(\mathrm{tag})}^{\mathrm{CM}}, \vec{p}^{~\mathrm{CM}}_{\tau(\mathrm{tag})} ]$ and $p_{\mathrm{track(tag)}}^{\mathrm{CM}}=[E_{\mathrm{track(tag)}}^{\mathrm{CM}},\vec{p}_{\mathrm{track(tag)}}^{~\mathrm{CM}}]$ are the four-momenta of tag-side tau and track in the CM frame. For the $\rho$ channel, photons in the tag side are considered in the calculation of the four-momentum of tag-side track.
Substituting $E_{\tau(\mathrm{tag})}^{\mathrm{CM}}=\sqrt{s}/2$, $m_{\tau(\mathrm{tag})}=m_{\tau}\sim 1.78$ GeV/$c^{2}$ and $m_{\mathrm{miss.track(tag)}}^{2} =  m_{\mathrm{miss.\ell\gamma.\mathrm{track(tag)}}}^{2}$,
where $m_{\mathrm{miss}.\ell\gamma.\mathrm{track(tag)}}^{2}$ is a missing mass squared of the event against the lepton and the photon in signal side and tag-side track,
\begin{equation} \label{eq:pvec}
\vec{p}_{\tau(\mathrm{tag})}^{~\mathrm{CM}}\cdot \vec{p}_{\mathrm{track(tag)}}^{~\mathrm{CM}} = \frac{m_{\mathrm{miss}.\ell\gamma.\mathrm{track(tag)}}^{2} - m_{\tau}^{2} - m_{\mathrm{track}}^{2} + \sqrt{s}E_{\mathrm{track(tag)}}^{\mathrm{CM}}}{2}.
\end{equation}
The $\xi_{\tau(\mathrm{tag}),\mathrm{track(tag)}}^{\mathrm{~CM}}$ is defined as
\begin{equation} \label{eq:xi}
\xi_{\tau(\mathrm{tag}),\mathrm{track(tag)}}^{\mathrm{~CM}} = \frac{\vec{p}_{\tau(\mathrm{tag})}^{\mathrm{~CM}}\cdot \vec{p}_{\mathrm{track}(\mathrm{tag})}^{\mathrm{~CM}}}{|\vec{p}_{\tau(\mathrm{tag})}^{\mathrm{~CM}}||\vec{p}_{\mathrm{track}(\mathrm{tag})}^{\mathrm{~CM}}|}.
\end{equation}  
Here, the momentum of the tag-side tau can be written as $\vec{p}_{\tau(\mathrm{tag})}^{\mathrm{~CM}}=-\vec{p}_{\tau({\mathrm{signal}})}^{\mathrm{~CM}}=-\vec{p}_{\gamma}^{\mathrm{~CM}}-\vec{p}_{\ell}^{\mathrm{~CM}}$ for signal events.
The $\xi$ variable corresponds to the cosine of the angle between the tau and the tag-side track, $\cos\theta_{\tau(\mathrm{tag}),\mathrm{track}(\mathrm{tag})}$ for an ideal signal event. 
Figure~\ref{fig:sr_cos_exp} shows the distribution of $\xi_{\tau(\mathrm{tag}),\mathrm{track(tag)}}^{\mathrm{~CM}}$.
The distribution for signal $\tau^{\pm}\rightarrow\ell^{\pm}\gamma$ events ranges from 0.0 to 1.0 except for detector resolution effect, whereas $\tau^{+}\tau^{-}$ background events have a broad distribution since Eq.~(\ref{eq:pvec}) is no more valid for background events.
Therefore, a criterion of $0.0<\xi_{\tau(\mathrm{tag}),\mathrm{track(tag)}}^{\mathrm{~CM}}<1.0$ is applied to suppress $\tau^{+}\tau^{-}$ background events.

\begin{figure}[htbp]
 \begin{minipage}{0.45\hsize}
  \begin{center}
   \includegraphics[width=70mm]{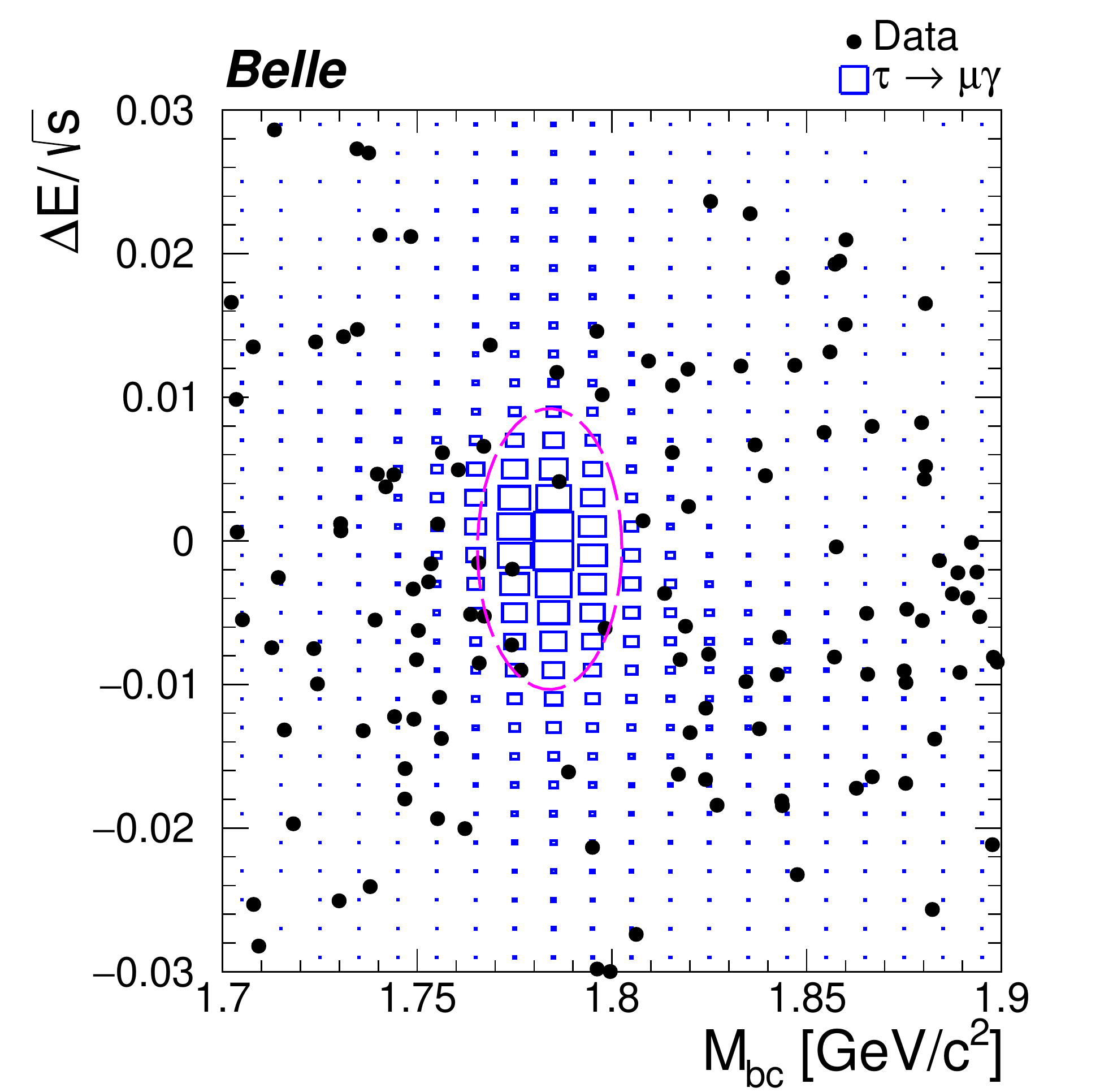}
   \hspace{1.5 cm} {(a) $\tau^{\pm}\rightarrow\mu^{\pm}\gamma$}
  \end{center}
 \end{minipage}
 \begin{minipage}{0.45\hsize}
  \begin{center}
   \includegraphics[width=70mm]{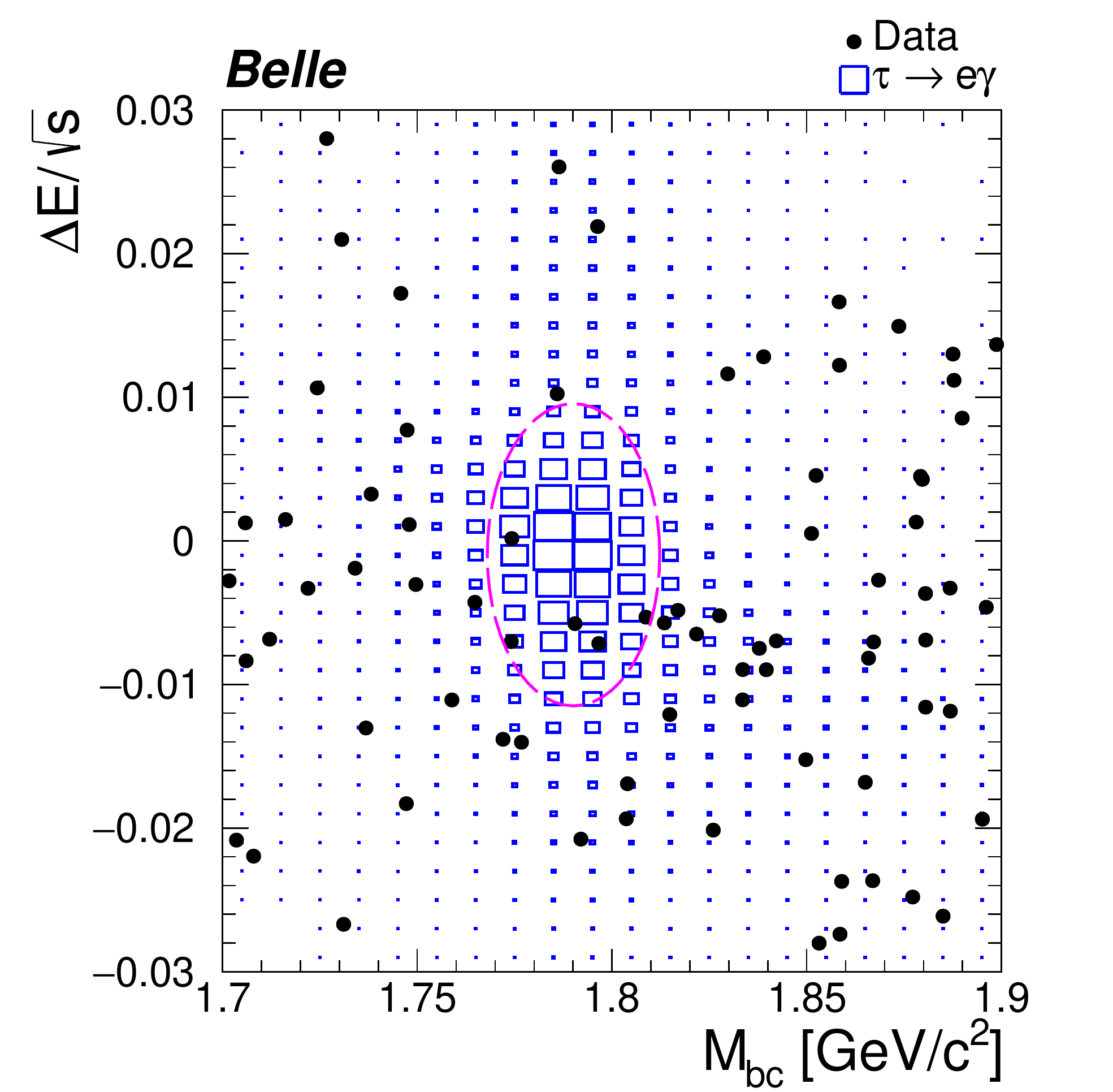}
   \hspace{1.5 cm} {(b) $\tau^{\pm}\rightarrow e^{\pm}\gamma$}
  \end{center}
 \end{minipage}
\caption{ \label{fig:sr_2d} 
Two-dimensional distributions of $\Delta E/\sqrt{s}$ vs. $M_{\mathrm{bc}}$ for (a)~$\tau^{\pm}\rightarrow\mu^{\pm}\gamma$ and (b)~$\tau^{\pm}\rightarrow e^{\pm}\gamma$ events.
Black points are data, blue squares are $\tau^{\pm}\rightarrow\ell^{\pm}\gamma$ signal MC events, and magenta ellipses show the signal region used in this analysis~($\pm 2\sigma$ region).
}
\end{figure}

The $\ell\gamma$ pair has an invariant mass of $m_{\tau}$ and the total energy in the CM frame of $E_{\ell\gamma}^{\mathrm{CM}} = \sqrt{s}/2$.
The signal region is defined by two kinematic variables: the beam-energy-constrained mass, $M_{\mathrm{bc}}$, and the normalized energy difference, $\Delta E/\sqrt{s}$, given as 
\begin{eqnarray}
  M_{\mathrm{bc}}= \sqrt{(E_{\mathrm{beam}}^{\mathrm{CM}})^{2} - |\vec{p}_{\ell\gamma}^{\mathrm{~CM}}|^{2}}, \\ \nonumber
\\ 
\Delta E/\sqrt{s} = ( E_{\ell\gamma}^{\mathrm{CM}} - \sqrt{s}/2 )/\sqrt{s},
\end{eqnarray}
where $E_{\mathrm{beam}}^{\mathrm{CM}}=\sqrt{s}/2$ and $\vec{p}_{\ell\gamma}^{\mathrm{~CM}}$ is the sum of the lepton and photon momenta in the CM frame.
Figure~\ref{fig:sr_2d} shows the two-dimensional distribution of $\Delta E/\sqrt{s}$ vs. $M_{\mathrm{bc}}$.
The signal events have $M_{\mathrm{bc}} \sim m_{\tau}$ and $\Delta E/\sqrt{s} \sim 0$ and in order to select them, an elliptical region around their expected values is adopted as follows:
\begin{eqnarray} \label{eq:sr}
  \frac{(M_{\mathrm{bc}} - \mu_{M_{\mathrm{bc}}})^{2}}{(2\sigma_{M_{\mathrm{bc}}})^{2}} &+& \frac{(\Delta E/\sqrt{s} - \mu_{\Delta E /\sqrt{s}})^{2}}{(2\sigma_{\Delta E/\sqrt{s}})^{2}} < 1.0, \\ \nonumber
  \\ \nonumber
  \sigma_{M_{\mathrm{bc}}} &=& 0.5(\sigma_{M_{\mathrm{bc}}}^{\mathrm{high}} + \sigma_{M_{\mathrm{bc}}}^{\mathrm{low}} ), \\ \nonumber
  \\ \nonumber
  \sigma_{\Delta E/\sqrt{s}} &=& 0.5(\sigma_{\Delta E/\sqrt{s}}^{\mathrm{high}} + \sigma_{\Delta E/\sqrt{s}}^{\mathrm{low}} ). \nonumber
\end{eqnarray}
Here, $\sigma^{\mathrm{high/low}}_{M_{\mathrm{bc}}}$ and $\sigma^{\mathrm{high/low}}_{\Delta E/\sqrt{s}}$ are the widths on the higher$/$lower side of the peak obtained by fitting the signal distribution to an asymmetric Gaussian function~\cite{HAYASAKA200816}.
The estimated resolutions are $\sigma_{M_{\mathrm{bc}}}^{\mathrm{high/low}} = 11.08\pm0.08/7.46\pm0.23$ MeV/$c^{2}$ and $\sigma_{\Delta E/ \sqrt{s}}^{\mathrm{high/low}} = (5.6\pm0.4)/(4.2\pm0.2)\times10^{-3}$ for $\tau^{\pm}\rightarrow\mu^{\pm}\gamma$ events, and $\sigma_{M_{\mathrm{bc}}}^{\mathrm{high/low}} = 11.55\pm0.27/10.59\pm0.19$ MeV/$c^{2}$ and $\sigma_{\Delta E/ \sqrt{s}}^{\mathrm{high/low}} = (6.1\pm0.7)/(4.4\pm0.3)\times10^{-3}$ for $\tau^{\pm}\rightarrow e^{\pm}\gamma$ events.
The mean values of the signal distributions are $\mu_{M_{\mathrm{bc}}} = 1.78$ MeV/$c^{2}$ and $\mu_{\Delta E/\sqrt{s}} = -0.6\times10^{-3}$ for $\tau^{\pm}\rightarrow\mu^{\pm}\gamma$ events, and $\mu_{M_{\mathrm{bc}}} = 1.79$ MeV/$c^{2}$ and $\mu_{\Delta E/\sqrt{s}} =  -1.0\times10^{-3}$ for $\tau^{\pm}\rightarrow e^{\pm}\gamma$ events.
The overall signal efficiency estimated using the above signal region is $3.7\%$ for $\tau^{\pm}\rightarrow\mu^{\pm}\gamma$ and $2.9\%$ for $\tau^{\pm}\rightarrow e^{\pm}\gamma$.

The most dominant background in the $\tau^{\pm}\rightarrow\mu^{\pm}\gamma$~($\tau^{\pm}\rightarrow e^{\pm}\gamma$) search arises from $\tau^{+}\tau^{-}$ events decaying to $\tau^{\pm}\rightarrow\mu^{\pm}\nu_{\mu}\nu_{\tau}$~($\tau^{\pm}\rightarrow e^{\pm}\nu_{e}\nu_{\tau}$) with a photon coming from initial-state radiation or beam background.
The $\mu^{+}\mu^{-}\gamma$ and $e^{+}e^{-}\gamma$ events are subdominant, with their contributions falling below $5\%$. Other backgrounds such as two-photon and $q\bar{q}$ are negligible in the signal region.

\section{Signal and background estimation}
To estimate the number of events in the signal region, we perform an unbinned maximum-likelihood fit with probability density functions~(PDFs) depending on $M_{\mathrm{bc}}$ and $\Delta E/\sqrt{s}$.
The likelihood function is defined in terms of the signal PDF~($S$), background PDF~($B$), and the number of signal events~($s$) and background events~($b$) as
\begin{equation} \label{eq:likelihood}
\mathcal{L} = \frac{\mathrm{e}^{-(s+b)}}{N!}\prod_{i=1}^{N}(sS_{i} + bB_{i}),
\end{equation}
where $N$ is the total number of observed events, $i$ denotes the event index, and $s$ and $b$ are the free parameters. The fit is performed to candidate events in the signal region defined by Eq.~(\ref{eq:sr}).
The signal PDF is obtained by smoothening the corresponding MC distribution and the background PDF uses the function described below.

Since the distributions of $M_{\mathrm{bc}}$ and $\Delta E/\sqrt{s}$ are well modeled for the $\tau^{+}\tau^{-}$ and $\mu^{+}\mu^{-}$ background events, the corresponding PDFs are determined using MC simulation.
The PDFs of $e^{+}e^{-}\gamma$ events are extracted from the data by applying an electron identification requirement, $\mathcal{L}_{e}>0.1$, to the track in the tag side. This is the same approach as in the previous publication~\cite{HAYASAKA200816}.
Since $M_{\mathrm{bc}}$ and $\Delta E/\sqrt{s}$ are almost independent of each other, the background PDF is written as
\begin{equation}
B(M_{\mathrm{bc}}, \Delta E/\sqrt{s}) = B(M_{\mathrm{bc}})\times B(\Delta E/\sqrt{s}).
\end{equation}
As the background events do not exhibit any peak and are rather flat in the $M_{\mathrm{bc}}$ distribution, a constant function is applicable to $f(M_{\mathrm{bc}})$.
In order to determine the $g(\Delta E/\sqrt{s})$ distribution, the requirement on $M_{\mathrm{bc}}$ is relaxed until enough statistics have been obtained. 
The background MC events with $M_{\mathrm{bc}}\in[1.74,~1.83]$~GeV/$c^{2}$ for $\tau^{+}\tau^{-}$ events and $M_{\mathrm{bc}}\in[1.60,~1.90]$~GeV/$c^{2}$ for $\mu^{+}\mu^{-}\gamma$ events are used in the case of $\tau^{\pm}\rightarrow\mu^{\pm}\gamma$ search.
For the $\tau^{\pm}\rightarrow e^{\pm}\gamma$ search, the background MC events with $M_{\mathrm{bc}}\in[1.70,~1.88]$~GeV/$c^{2}$ for $\tau^{+}\tau^{-}$ events and $M_{\mathrm{bc}}\in[1.73,~1.85]$~GeV/$c^{2}$ for $e^{+}e^{-}\gamma$ events are used.
The $\Delta E/\sqrt{s}$ distribution for $\tau^{+}\tau^{-}$ background is described by a sum of Landau and exponential functions for both $\tau^{\pm}\rightarrow\mu^{\pm}\gamma$ and $\tau^{\pm}\rightarrow e^{\pm}\gamma$ searches.
The distribution for $\mu^{+}\mu^{-}\gamma$ and $e^{+}e^{-}\gamma$ is described by a sum of Landau and Gaussian functions~\cite{HAYASAKA200816}.

\begin{figure}[htbp]
 \begin{minipage}{0.45\hsize}
  \begin{center}
   \includegraphics[width=70mm]{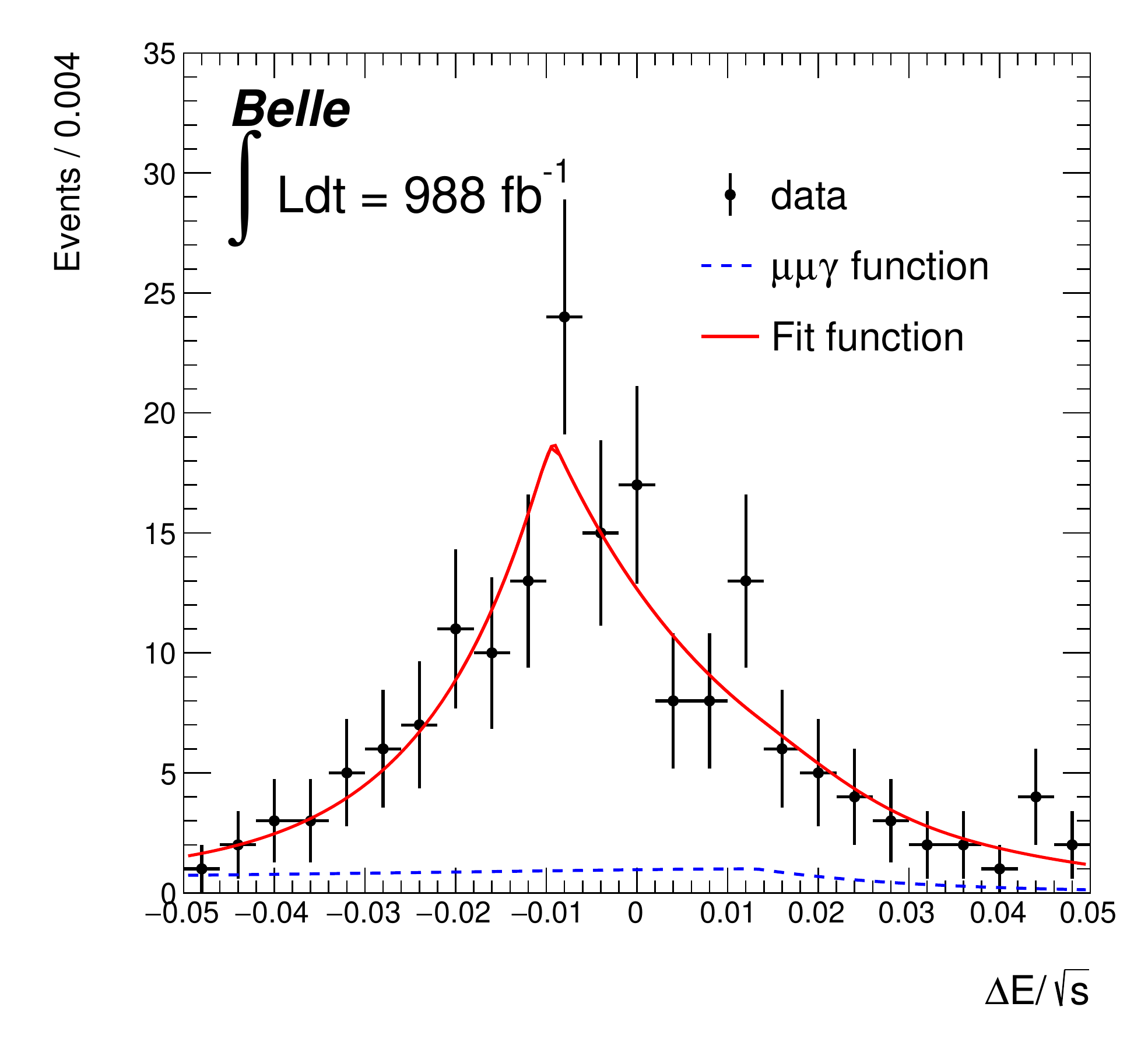}
   \hspace{1.5 cm} {(a) $\tau^{\pm}\rightarrow\mu^{\pm}\gamma$}
  \end{center}
 \end{minipage}
 \begin{minipage}{0.45\hsize}
  \begin{center}
   \includegraphics[width=70mm]{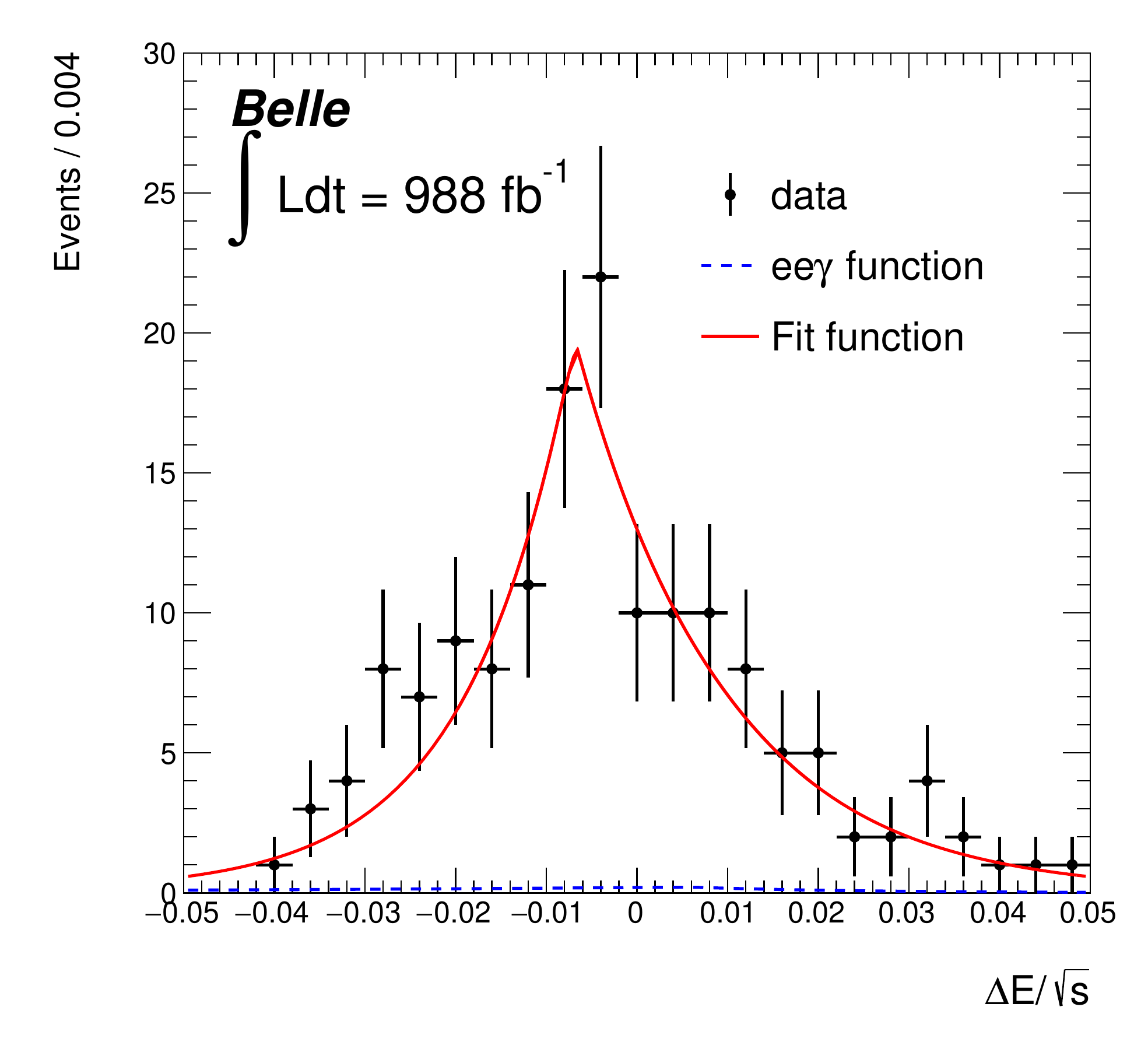}
   \hspace{1.5 cm} {(b) $\tau^{\pm}\rightarrow e^{\pm}\gamma$}
  \end{center}
 \end{minipage}
\caption{ \label{fig:sideband_fit} 
  $\Delta E/\sqrt{s}$ distribution in the sideband.
  The black points with error bars are the data and red curves show the fit result of the background PDF.
}
\end{figure}

The total background PDFs~($B^{\mathrm{tot}}_{0}, B^{\mathrm{tot}}_{1}$) are obtained by combining each background function:
\begin{eqnarray}
  B^{\mathrm{tot}}_{0} &=& C_{0}B_{\tau\tau} + C_{1}B_{\mu\mu\gamma}, \\ \nonumber
  \nonumber \\
  B^{\mathrm{tot}}_{1} &=& C_{2}B_{\tau\tau} + C_{3}B_{ee\gamma}, \\ \nonumber
\end{eqnarray}  
where $B_{\tau\tau}$, $B_{\mu\mu\gamma}$, and $B_{ee\gamma}$ are the PDFs for $\tau^{+}\tau^{-}$, $\mu^{+}\mu^{-}$, and $e^{+}e^{-}\gamma$ background events, and $C_{0}$ to $C_{3}$ are the free parameters determined by a fit.
The fit is performed to the sideband data defined as $M_{\mathrm{bc}}\in[1.60,~1.74]\cup[1.83,~1.97]$ GeV/$c^{2}$ for the $\tau^{\pm}\rightarrow\mu^{\pm}\gamma$ search and $M_{\mathrm{bc}}\in[1.57,~1.75]\cup[1.85,~2.00]$ GeV/$c^{2}$ for the $\tau^{\pm}\rightarrow e^{\pm}\gamma$ search.
Figure~\ref{fig:sideband_fit} shows $\Delta E/\sqrt{s}$ distributions in the sideband.
After performing the fit, we obtain $C_{0}=19.3\pm1.8$, $C_{1}=1.0\pm0.7$ for the $\tau^{\pm}\rightarrow\mu^{\pm}\gamma$ search, and $C_{2}=19.7\pm1.9$, $C_{3}=0.2\pm0.7$ for the $\tau^{\pm}\rightarrow e^{\pm}\gamma$ search.
The $\tau^{+}\tau^{-}$ background events are dominant for both search channels and consistent with the MC expectation.
The expected number of background events is $5.8\pm0.4$ for the $\tau^{\pm}\rightarrow\mu^{\pm}\gamma$ search and $5.1\pm0.4$ for the $\tau^{\pm}\rightarrow e^{\pm}\gamma$ search.

The total number of observed events is 5 in both the $\tau^{\pm}\rightarrow\mu^{\pm}\gamma$ and $\tau^{\pm}\rightarrow e^{\pm}\gamma$ searches, as shown in Figure~\ref{fig:sr_2d}.
By using the aforementioned signal and background PDFs, we perform the likelihood fit defined in Eq.~(\ref{eq:likelihood}).
The results of the likelihood fit are $s=-0.3^{+1.8}_{-1.3}$, $b=5.3^{+3.2}_{-2.3}$ for $\tau^{\pm}\rightarrow\mu^{\pm}\gamma$, and $s=-0.5^{+4.4}_{-3.6}$, $b=5.5^{+5.2}_{-4.1}$ for $\tau^{\pm}\rightarrow e^{\pm}\gamma$.

We estimate the systematic uncertainties associated with track and photon reconstruction efficiencies, photon energy calibration, luminosity, trigger efficiencies, and background PDF modeling.
A summary of these systematic uncertainties is given in Table~\ref{tab:syst_all}.

The uncertainty in track reconstruction efficiencies is estimated with partially reconstructed $D^{*+}\rightarrow D^{0}\pi^{+}$, $D^{0}\rightarrow K^{0}_{S}\pi^{+}\pi^{-}$ events.
The systematic uncertainty of $0.35\%$ is assigned per track, and thus a total uncertainty of $0.7\%$ is estimated for our analysis.
The efficiencies of photon reconstruction are estimated with radiative Bhabha events. The efficiencies in MC simulation agree with that in data, and the associated uncertainty is $2.0\%$.
As discussed earlier, the uncertainty due to photon energy calibration is estimated with $e^{+}e^{-}\rightarrow\mu^{+}\mu^{-}\gamma$ events, and amounts to $3.2\%$.
The uncertainty in the integrated luminosity is $1.4\%$. 
The trigger efficiencies are evaluated by comparing the data sideband and MC simulation, and estimated to be $2.1\%$ for $\tau^{\pm}\rightarrow\mu^{\pm}\gamma$ and $3.4\%$ for $\tau^{\pm}\rightarrow e^{\pm}\gamma$ analysis.
These are the uncertainties related to overall signal efficiency.
The uncertainty due to background PDF modeling is evaluated by varying the fixed PDF parameters.
By changing each of the fixed parameters by $\pm1\sigma$, the number of signal events obtained from the fit is checked, and the relative difference from the nominal value is assigned as the systematic uncertainty.
The estimated uncertainty is $3.3\%$ for $\tau^{\pm}\rightarrow\mu^{\pm}\gamma$ and $3.7\%$ for $\tau^{\pm}\rightarrow e^{\pm}\gamma$.
The uncertainties due to limited MC statistics and particle identification are negligible compared to the other uncertainties described above.

\begin{table}[htbp]
\begin{center}
\caption{\label{tab:syst_all} Systematic uncertainties~(in $\%$) considered in this analysis.}
\vspace{0.2cm}
\begin{tabular}{l @{\hspace{1cm}} |c @{\hspace{1cm}} c} \hline
Source                   & $\tau^{\pm}\rightarrow\mu^{\pm}\gamma$   & $\tau^{\pm}\rightarrow e^{\pm}\gamma$  \\ \hline
Track reconstruction efficiency         & 0.7  & 0.7  \\ 
Photon reconstruction efficiency        & 2.0  & 2.0  \\ 
Photon energy calibration               & 3.2  & 3.2  \\ 
Integrated luminosity                   & 1.4  & 1.4  \\ 
Trigger efficiency                      & 2.1  & 3.4  \\ 
Background PDF modeling                 & 3.3  & 3.7  \\ \hline
\end{tabular}
\end{center}
\end{table}

\section{Result}
Since no significant excess of the signal events is observed in data, the upper limits at the $90\%$ confidence level~(CL) are evaluated using toy MC simulations.
We generate toy signal and background events based on their PDFs while fixing the number of background events~($\tilde{b}$) and varying the number of signal events~($\tilde{s}$).
For every assumed $\tilde{s}$, 10,000 pseudoexperiments are generated following Poisson statistics with the means $\tilde{s}$ and $\tilde{b}$ for signal and background, respectively.
In order to obtain the expected~(observed) upper limits on the branching fraction at $90\%$ CL, the $\tilde{s}$ value that gives a 90\% probability for $\tilde{s}$ larger than zero~(fitted signal yield) is taken: $\tilde{s}_{90}$.
The method to incorporate the systematic uncertainties into a branching fraction discussed in Ref.~\cite{cleo} is adopted in this analysis: the uncertainties related to overall signal efficiency and background PDF modeling are treated separately.
The likelihood defined in Eq.~(\ref{eq:likelihood}) is convolved with a Gaussian function of width equal to the systematic uncertainty, so the $\tilde{s}$ and $\tilde{b}$ values are smeared accordingly.
The uncertainties inflate the upper limits on the branching fraction by $\sim$2-3\%; this effect is not large and consistent with the past results~\cite{HAYASAKA200816}.
The expected upper limits on the branching fraction $\mathcal{B}(\tau^{\pm}\rightarrow\ell^{\pm}\gamma)$ at $90\%$ CL is calculated as $\mathcal{B}(\tau^{\pm}\rightarrow\mu^{\pm}\gamma)<4.9 \times 10^{-8}$ and $\mathcal{B}(\tau^{\pm}\rightarrow e^{\pm}\gamma)<6.4 \times 10^{-8}$.
Our expected limits are 1.6--1.8 times more stringent compared to the previous Belle results~\cite{HAYASAKA200816}.

The toy MC simulation provides an observed upper limit on signal at the $90\%$ CL as $\tilde{s}_{90}=2.8$~($\tilde{s}_{90}=3.0$) events from the fit for $\tau^{\pm}\rightarrow\mu^{\pm}\gamma$~($\tau^{\pm}\rightarrow e^{\pm}\gamma$).
The observed upper limits on the branching fractions are
\begin{eqnarray}
  \mathcal{B}(\tau^{\pm}\rightarrow\mu^{\pm}\gamma) &<& \frac{\tilde{s}_{90}}{2 \epsilon N_{\tau\tau}} = 4.2 \times 10^{-8}, \\
  \mathcal{B}(\tau^{\pm}\rightarrow e^{\pm}\gamma) &<& \frac{\tilde{s}_{90}}{2 \epsilon N_{\tau\tau}} = 5.6 \times 10^{-8},
\end{eqnarray}  
where $N_{\tau\tau}=(912\pm14)\times10^{6}$, and the signal efficiencies are $\epsilon=3.7\%$ and $2.9\%$ for $\tau^{\pm}\rightarrow\mu^{\pm}\gamma$ and $\tau^{\pm}\rightarrow e^{\pm}\gamma$, respectively.

\section{Summary}
In this paper, a search conducted for the charged-lepton-flavor-violating decays, $\tau^{\pm}\rightarrow\mu^{\pm}\gamma$ and $\tau^{\pm}\rightarrow e^{\pm}\gamma$, at the Belle experiment is reported.
It uses 988~fb$^{-1}$ of data, about twice the size used in the previous Belle analysis~\cite{HAYASAKA200816}.
In addition, requirements with new observables of energy asymmetry and beam-energy-constrained mass are introduced to further reduce background events. The selection is optimized by taking into account the different tag-side modes to maximize search sensitivities.
Lastly, the photon energy is calibrated using radiative muon events.
Thanks to those improvements and $1.9$ times data, our expected limits are 1.6--1.8 times more stringent compared to the previous Belle results~\cite{HAYASAKA200816}.
With the absence of signal in any modes, the upper limits are set on branching fractions: $\mathcal{B}(\tau^{\pm}\rightarrow\mu^{\pm}\gamma) <  4.2 \times 10^{-8}$ and $\mathcal{B}(\tau^{\pm}\rightarrow e^{\pm}\gamma) <  5.6 \times 10^{-8}$ at the $90\%$ confidence level. The observed limit on the $\tau^{\pm}\rightarrow\mu^{\pm}\gamma$ decay is the most stringent to date.


\acknowledgments
We thank the KEKB group for the excellent operation of the
accelerator; the KEK cryogenics group for the efficient
operation of the solenoid; and the KEK computer group, and the Pacific Northwest National
Laboratory (PNNL) Environmental Molecular Sciences Laboratory (EMSL)
computing group for strong computing support; and the National
Institute of Informatics, and Science Information NETwork 5 (SINET5) for
valuable network support.  We acknowledge support from
the Ministry of Education, Culture, Sports, Science, and
Technology (MEXT) of Japan, the Japan Society for the 
Promotion of Science (JSPS) including in particular the Grant-in-Aid for Scientific Research (A) 19H00682, and the Tau-Lepton Physics 
Research Center of Nagoya University; 
the Australian Research Council including grants
DP180102629, 
DP170102389, 
DP170102204, 
DP150103061, 
FT130100303; 
Austrian Federal Ministry of Education, Science and Research (FWF) and
FWF Austrian Science Fund No.~P~31361-N36;
the National Natural Science Foundation of China under Contracts
No.~11435013,  
No.~11475187,  
No.~11521505,  
No.~11575017,  
No.~11675166,  
No.~11705209;  
Key Research Program of Frontier Sciences, Chinese Academy of Sciences (CAS), Grant No.~QYZDJ-SSW-SLH011; 
the  CAS Center for Excellence in Particle Physics (CCEPP); 
the Shanghai Pujiang Program under Grant No.~18PJ1401000;  
the Shanghai Science and Technology Committee (STCSM) under Grant No.~19ZR1403000; 
the Ministry of Education, Youth and Sports of the Czech
Republic under Contract No.~LTT17020;
Horizon 2020 ERC Advanced Grant No.~884719 and ERC Starting Grant No.~947006 ``InterLeptons'' (European Union);
the Carl Zeiss Foundation, the Deutsche Forschungsgemeinschaft, the
Excellence Cluster Universe, and the VolkswagenStiftung;
the Department of Atomic Energy (Project Identification No. RTI 4002) and the Department of Science and Technology of India; 
the Istituto Nazionale di Fisica Nucleare of Italy; 
National Research Foundation (NRF) of Korea Grant
Nos.~2016R1\-D1A1B\-01010135, 2016R1\-D1A1B\-02012900, 2018R1\-A2B\-3003643,
2018R1\-A6A1A\-06024970, 2018R1\-D1A1B\-07047294, 2019K1\-A3A7A\-09033840,
2019R1\-I1A3A\-01058933;
Radiation Science Research Institute, Foreign Large-size Research Facility Application Supporting project, the Global Science Experimental Data Hub Center of the Korea Institute of Science and Technology Information and KREONET/GLORIAD;
the Polish Ministry of Science and Higher Education and 
the National Science Center;
the Ministry of Science and Higher Education of the Russian Federation, Agreement 14.W03.31.0026, 
and the HSE University Basic Research Program, Moscow; 
University of Tabuk research grants
S-1440-0321, S-0256-1438, and S-0280-1439 (Saudi Arabia);
the Slovenian Research Agency Grant Nos. J1-9124 and P1-0135;
Ikerbasque, Basque Foundation for Science, Spain;
the Swiss National Science Foundation; 
the Ministry of Education and the Ministry of Science and Technology of Taiwan;
and the United States Department of Energy and the National Science Foundation.




\begin{thebibliography}{99}

\bibitem{SMnu}
  X.-Y. Pham, Eur. Phys. J. C  {\bf 8}, 3 (1999)

\bibitem{MSSM}
  A. Brignole and A. Rossi, Nucl. Phys. B {\bf 701}, 3 (2004).

\bibitem{GUT}
  L. Calibbi, A. Faccia, A. Masiero, and S.K. Vempati, Phys. Rev. D {\bf 74}, 116002 (2006).
  
\bibitem{SeeSaw}
  J.R. Ellis, J. Hisano, M. Raidal, and Y. Shimizu, Phys. Rev. D {\bf 66}, 115013 (2002).

\bibitem{HAYASAKA200816}
  K. Hayasaka {\it et al.}~(Belle Collaboration), Phys. Lett. B {\bf 666}, 16 (2008).

\bibitem{BhaBhaResult}
  B. Aubert {\it et al.}~(BaBar Collaboration), Phys. Rev. Lett. {\bf 104}, 021802 (2010).

\bibitem{KEKB}
  S. Kurokawa and E.~Kikutani, Nucl. Instrum. Meth. Phys. Res., Sect. A {\bf 499}, 1 (2003), and other papers included in this Volume; T.~Abe {\it et al.}, Prog. Theor. Exp. Phys. {\bf 2013}, 03A001 (2013) and references therein.
  
\bibitem{PEP2}
  J. P. Lees {\it et al.}~(BABAR Collaboration), Nucl. Instrum. Meth. A {\bf 726}, 203 (2013).

\bibitem{Luminosity}
  J. Brodzicka {\it et al.}~(Belle Collaboration), Prog. Theor. Exp. Phys. {\bf 2012}, 2050 (2012).

\bibitem{Belle}
  A. Abashian {\it et al.} (Belle Collaboration), Nucl. Instrum. Meth. Phys. Res., Sect. A {\bf 479}, 117 (2002).

\bibitem{KKMC}
  S. Jadach, B. F. L. Ward, and Z. Was, Comput. Phys. Commun. {\bf 130}, 260 (2000).

\bibitem{BHLUMI}
  S. Jadach {\it et al.}, Comp. Phys. Commun. {\bf 70}, 305 (1992).

\bibitem{AAFHB}
  F. A. Berends {\it et al.}, Comp. Phys. Commun. {\bf 40}, 285 (1986).

\bibitem{EvtGen}
  D. J. Lange, Nucl. Instrum. Meth. Phys. Res., Sect. A {\bf 462}, 152 (2001).

\bibitem{GEANT3}
  R. Brun {\it et al.}, GEANT 3.21, CERN Report DD/EE/84-1 (1984).

\bibitem{TestBeam}
  H. Ikeda {\it el al.}, Nucl. Instrum. Meth. Phys. Res., Sect. A {\bf 441}, 401 (2000).

\bibitem{Muon}
  A. Abashian {\it et al.}, Nucl. Instrum. Meth. Phys. Res., Sect. A {\bf 491}, 69 (2002).

\bibitem{Electron}
  K. Hanagaki {\it et al.}, Nucl. Instrum. Meth. Phys. Res., Sect. A {\bf 485}, 490 (2002).

\bibitem{thrust}
  S. Brandt, C. Peyrou, R. Sosnowski, and A. Wroblewski, Phys. Lett. {\bf 12}, 57 (1964); E. Farhi, Phys. Rev. Lett. {\bf 39}, 1587 (1977).

\bibitem{cleo}
  S. Ahmed {\it et al.} (CLEO Collaboration), Phys. Rev. D {\bf 61}, 071101 (2000).





\end{thebibliography}
\end{document}